\documentclass[prd, 10pt,aps,nofootinbib, floatfix, notitlepage,twocolumn,longbibliography]{revtex4-1}

\usepackage{amsmath,amsfonts,amssymb}
\usepackage{mathrsfs}
\usepackage{graphicx}
\usepackage[english]{babel} 
\usepackage{url} 
\usepackage{color}
\usepackage{bbold}
\usepackage{adjustbox}
 \usepackage[utf8]{inputenc} 
\usepackage[colorlinks=true,citecolor=blue,urlcolor=blue]{hyperref}

\newcommand{\be}{\begin{equation}}
\newcommand{\ee}{\end{equation}}
\newcommand{\bes}{\begin{equation*}}
\newcommand{\ees}{\end{equation*}}

\newcommand{\LCDM}{$\Lambda$CDM}
\begin{document}

\title{Early Dark Energy Can Resolve The Hubble Tension}
\author{Vivian Poulin$^1$}
\author{Tristan L.~Smith$^2$}
\author{Tanvi Karwal$^1$}
\author{Marc Kamionkowski$^1$}

\affiliation{$^1$Department of Physics and Astronomy, Johns
				Hopkins University, 3400 N.\ Charles St., Baltimore, MD 21218, United States}
\affiliation{$^2$Department of Physics and Astronomy, Swarthmore College, 500 College Ave., Swarthmore, PA 19081, United States}

\begin{abstract}
Early dark energy (EDE) that behaves like a cosmological constant at early times (redshifts $z\gtrsim3000$) and then dilutes away like radiation or faster at later times can solve the Hubble tension.  In these models, the sound horizon at decoupling is reduced resulting in a larger value of the Hubble parameter $H_0$ inferred from the cosmic microwave background (CMB). We consider two physical models for this EDE, one involving an oscillating scalar field and another a slowly-rolling field.  We perform a detailed calculation of the evolution of perturbations in these models.  A Markov Chain Monte Carlo search of the parameter space for the EDE parameters, in conjunction with the standard cosmological parameters, identifies regions in which $H_0$ inferred from {\it Planck} CMB data agrees with the SH0ES local measurement.  In these cosmologies, current baryon acoustic oscillation and supernova data are described as successfully as in \LCDM, while the fit to {\it Planck} data is slightly improved. Future CMB and large-scale-structure surveys will further probe this scenario.
\end{abstract}

\date{\today}

\maketitle
 

Local measurements of the Hubble parameter, from supernovae
\cite{Riess:2016jrr,Riess:2018byc} and lensing time delays
\cite{2017MNRAS.465.4914B,Birrer:2018vtm}, disagree with the value inferred from a \LCDM~fit to
the cosmic microwave background (CMB) \cite{Ade:2015xua,Aghanim:2018eyx}, with local
measurements suggesting a higher value.  This discrepancy is not
easily explained by any obvious systematic effect in either
measurement \cite{Efstathiou:2013via,Addison:2015wyg,Aghanim:2016sns,Aylor:2018drw}, and so
increasing attention is focusing on the possibility that this
``Hubble tension'' may be indicating new physics beyond the
standard \LCDM~cosmological model \cite{Freedman:2017yms,Feeney:2017sgx}.

However, theoretical explanations for the Hubble tension are not
easy to come by.  The biggest challenge remains the very
precisely determined angular scale of the acoustic peaks in the
CMB power spectrum, which fix the ratio of the sound horizon at decoupling to the  distance to the CMB surface of last scatter.  
Possible late-time resolutions include a phantom-like dark energy (DE) component \cite{DiValentino:2016hlg,DiValentino:2017zyq}, a vacuum phase transition \cite{DiValentino:2017rcr,Khosravi:2017hfi,Banihashemi:2018has,Banihashemi:2018oxo}, or interacting DE \cite{Kumar:2016zpg,DiValentino:2017iww}. However, these resolutions are tightly constrained \cite{Riess:2016jrr,DiValentino:2017zyq,Addison:2017fdm,DiValentino:2017iww} by late-time observables, especially those from baryon acoustic oscillations (BAO) \cite{Beutler:2011hx,Ross:2014qpa,Alam:2016hwk}. Model-independent parameterizations of the late-time expansion history are similarly constrained \cite{Bernal:2016gxb,Zhao:2017cud,Poulin:2018zxs}.  An early-time resolution, which reduces the sound horizon with additional radiation energy density \cite{Riess:2016jrr,Riess:2018byc}, is constrained by BAO and by the higher peaks in the CMB power spectrum \cite{Bernal:2016gxb,DiValentino:2017iww}. It is also possible to address the Hubble tension through a modification of gravity \cite{Barreira:2014jha,Umilta:2015cta,
Ballardini:2016cvy,Renk:2017rzu,Belgacem:2017cqo,Nunes:2018xbm,Lin:2018nxe}.

Another early-time resolution~\cite{Karwal:2016vyq,Mortsell:2018mfj} is an exotic
early dark energy (EDE) that behaves like a cosmological constant
before some critical redshift $z_c$ but whose energy density then dilutes faster than radiation.  This addresses the Hubble tension by increasing the early expansion rate 
while leaving the later evolution of the Universe
unchanged.  Ref.~\cite{Karwal:2016vyq} investigated the effects on the
CMB under the 
assumption that the dark energy exhibited no spatial
fluctuations.  A simple Fisher analysis of CMB data suggested that the model could push the CMB-inferred $H_0$ in the right direction, but not enough.

Here, we present two physical models for EDE, one that involves an oscillating scalar field and another with a slowly-rolling scalar field.  These models allow us to perform a complete analysis of the growth of perturbations and of CMB fluctuations.  We then perform a thorough search of the parameter space for the scalar-field model parameters, along with the classical cosmological parameters.  Doing so, we find regions of the combined parameter space where the CMB likelihoods match (and even slightly improve upon) those in the best-fit \LCDM\ model with values of $H_0$ consistent with those from local measurements.  
Moreover, our cosmological model is in good agreement with constraints from BAO \cite{Beutler:2011hx,Ross:2014qpa,
Alam:2016hwk} and the Pantheon supernovae dataset \citep{2018ApJ...859..101S}.  The fact that both an oscillating and slowly-rolling scalar field can resolve the Hubble tension indicates further that the success of the resolution does not depend on the detailed mechanism that underlies it.
Our resolution requires a $\sim5\%$ contribution from EDE to the total energy density at redshift $z\simeq5000$ that then dilutes later.  Interestingly, hints for such an increased expansion rate and/or reduced sound horizon had been previously identified \cite{Hojjati:2013oya,Aylor:2018drw}.

Our first model for EDE is nominally a scalar field $\varphi$ with a potential $V(\varphi) \propto (1-\cos[\varphi/f])^n$ \cite{Kamionkowski:2014zda}.  At early times, the field is frozen and acts as a cosmological constant, but when the Hubble parameter drops below some value, at a critical redshift $z_c =a_c^{-1}-1$, the field begins to oscillate and then behaves as a fluid with an equation of state $w_n= (n-1)/(n+1)$.  In practice, numerical evolution of the scalar-field equations of motion becomes extremely difficult once the oscillations become rapid compared with the expansion rate, and so our numerical work is accomplished with an effective-fluid approach \cite{Poulin:2018dzj} that has been tailored specifically for this potential.  Still, as that work (and discussion below) indicates, our conclusions do not depend on the details of the potential and would work just as well with, e.g., a simpler $\varphi^{2n}$ potential.  Our second model is a field that slowly rolls down a potential that is linear in $\varphi$ at early times and asymptotes to zero at late times.  Numerical evolution of the scalar-field equations of motion confirm that the resolutions we find here with the effective-fluid approach are valid for that model as well; details will be presented elsewhere \cite{Tanvi}. 
  
In the effective-fluid approximation, the EDE energy density evolves as \cite{Poulin:2018dzj}
\begin{equation}
 \Omega_{\varphi}(a)= \frac{2 \Omega _{\varphi}(a_c)}{\left(a/a_c\right)^{3( w_n+1)}+1},\label{eq:omegaFit}
 \end{equation}
 which has an associated equation-of-state parameter
 \begin{eqnarray}
 w_{\varphi}(z)& = & \frac{1+w_n}{1+(a_c/a)^{3(1+w_n)}}-1.    \label{eq:wphi}
 \end{eqnarray}
It asymptotically approaches $-1$ as $a \rightarrow 0$ and $w_n$ for $a \gg a_c$, showing that the energy density is constant at early times and dilutes as $a^{-3(1+w_n)}$ once the field is dynamical \cite{Turner:2001mx}. The homogeneous EDE energy density dilutes like matter for $n=1$, like radiation for $n=2$ and faster than radiation whenever $n\geq3$. 
For $n\to\infty$, on reaching the minimum of the potential, $w_{\infty}=1$ (i.e. the scalar field is fully dominated by its kinetic energy) and the energy density dilutes as $a^{-6}$.

The equations governing the evolution of the perturbations to the effective density $\delta_{\varphi}$ and heat flux $u_\varphi \equiv (1+w_\varphi) \theta_\varphi$, where $\theta_\varphi$ is the bulk velocity perturbation,\footnote{It is known \citep{Hlozek:2014lca,Poulin:2018dzj} that for a scalar field the evolution equation of the velocity perturbation is unstable as $w\rightarrow -1$ and we therefore solve for the heat-flux.} can be written as discussed in Refs.~\citep{Hu:1998kj,Hlozek:2014lca,Poulin:2018dzj}. 
Solving these equations requires the specification of the EDE equation-of-state $w_{\varphi}(z)$, the adiabatic sound speed $c_{\rm a}^2 \equiv \dot P_{\varphi}/\dot \rho_{{\varphi}}$
and effective sound speed $c_s^2\equiv \delta p_\varphi/\delta \rho_\varphi$ (defined in the rest-frame of the field). During slow roll and assuming $\dot{\varphi}_i=0$, generic scalar fields have $w_\varphi \simeq -1$, $c_{\rm a}^2 \simeq -7/3$, and $c_s^2 =1$ \cite{Hlozek:2014lca,Poulin:2018dzj}. When the field becomes dynamical, $w_a$ and $c_{\rm a}^2$ can be calculated from the background parametrization. The exact behavior of $c_s^2$ depends on the particular shape of the potential as described in Ref.~\cite{Poulin:2018dzj}. 
We also note that, just as with the background dynamics, this parametrization describes the case of the slow-roll model \cite{Tanvi} by taking the limit $n\to\infty$ and setting $c_s^2 = 1$ \citep{Hu:1998kj}. 

We run a Markov Chain Monte Carlo (MCMC) using the public code {\sc MontePython-v3}\footnote{\url{https://github.com/brinckmann/montepython_public}} \citep{Audren:2012wb,Brinckmann:2018cvx} and a modified version of the {\sc CLASS}-code \cite{Lesgourgues:2011re,
Blas:2011rf}. 
We perform the analysis with a Metropolis-Hasting algorithm, assuming flat priors on $\{\omega_b,\omega_{\rm cdm},\theta_s,A_s,n_s,\tau_{\rm reio},\Omega_{\varphi,0} ,{\rm Log}_{10}(a_c),\phi_i\}$. In addition, we run separate MCMCs to compare\footnote{The $n=1$ case leads to an over-production of cdm once the field starts diluting. We checked explicitly that it does not solve the $H_0$-tension by performing a dedicated run.} $n = (2,3,\infty)$. Following the {\em Planck} collaboration, we model free-streaming neutrinos as two massless species and one massive with $M_\nu=0.06$ eV \cite{Ade:2018sbj}. 
Our data sets include the latest SH0ES measurement of the present-day Hubble rate $H_0=73.52\pm1.62$ km/s/Mpc~\cite{Riess:2018byc}, {\em Planck} high-$\ell$ and low-$\ell$ TT,TE,EE and lensing likelihood \cite{Aghanim:2015xee}.
We also include BAO measurements from 6dFGS at $z = 0.106$~\cite{Beutler:2011hx}, from the MGS galaxy sample of SDSS at $z = 0.15$~\cite{Ross:2014qpa}, and
  from the CMASS and LOWZ galaxy samples of BOSS DR12 at $z = 0.38$, $0.51$, and $0.61$~\cite{Alam:2016hwk}. 
Note that the BOSS DR12 measurements also include measurements of the growth function $f\sigma_8(z)$.  Additionally, we use the Pantheon\footnote{\url{https://github.com/dscolnic/Pantheon}} supernovae dataset \cite{2018ApJ...859..101S}, which includes measurements of the luminosity distances of 1048 SNe Ia in the redshift range $0.01 < z < 2.3$. Moreover, there are many nuisance parameters that we analyze together with the cosmological ones using a Choleski decomposition \citep{Lewis:2013hha}. We consider chains to be converged using the Gelman-Rubin \citep{Gelman:1992zz} criterion $R -1<0.1$. 

\begin{figure}[!h]
\centering
\includegraphics[scale=0.43]{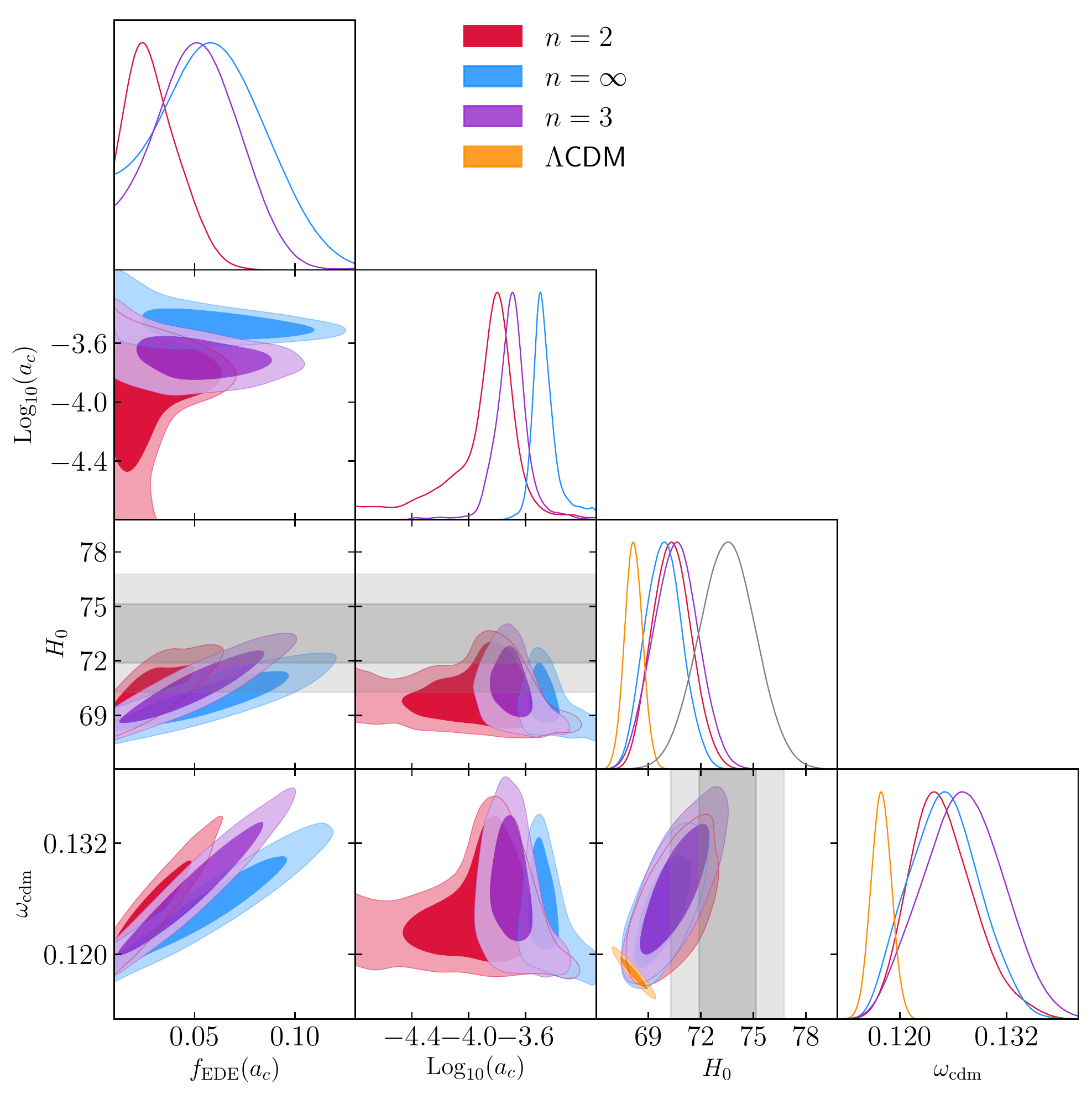}
\caption{Comparison between the marginalized 1D and 2D posterior distributions of $H_0$, $\omega_{\rm cdm}$, $f_{\rm EDE}(a_c)$ and Log$_{10}(a_c)$ in the EDE cosmology with $n=2$, $n=3$ and $n=\infty$. The best fit value of $H_0$ in \LCDM~is shown in orange; the one from SH0ES is shown in grey. 
\label{fig:posteriors}}
\end{figure}

In Fig.~\ref{fig:posteriors}, we show the marginalized 1D and 2D posterior distributions of $H_0$, $\omega_{\rm cdm}$, $f_{\rm EDE}(a_c)$ and Log$_{10}(a_c)$ in $\Lambda$CDM and in the EDE cosmology with $n=2$, $3$ and $n\to \infty$, where $f_{\rm EDE}(a_c) \equiv \Omega_{\varphi}(a_c)/\Omega_{\rm tot}(a_c)$. We report the best-fit $\chi^2$ for each experiment in Table \ref{table:chi2_preliminary}, while  the reconstructed mean, best fit and 1$\sigma$ confidence interval of the cosmological parameters are given in Table~\ref{table:param_values}.
We find that the best-fit $\chi^2$ in the EDE cosmology is reduced by $-9$ to $-14$ compared to \LCDM\ using the same collection of data-sets.
This reduction in the $\chi^2$ is not only driven by an improved fit of SH0ES data, but also by an improved fit of CMB data compared to a $\Lambda$CDM fit to all data-sets. 
Interestingly, in the global fit, the EDE fits {\it Planck} data slightly better than  \LCDM\ fitted on {\em Planck} {\em only}\footnote{ The fit of $\Lambda$CDM on  {\em Planck} only yields $\chi^2_{Planck}\simeq 12951.5$ for the exact same precision parameters as the one used in the EDE fits and convergence criterion $R-1 < 0.008$. It can vary slightly from the one quoted in {\em Planck} tables \cite{Ade:2015xua}. }. 
This is in stark contrast with the case of extra-relativistic degrees of freedom, for which the $\chi^2$ of CMB and BAO data degrades (as shown on the last column of Table~\ref{table:chi2_preliminary} and also found by Refs.~\cite{Bernal:2016gxb,DiValentino:2016hlg,Poulin:2018zxs}). 
In order to get an estimate of the statistical preference of the EDE cosmology compared to $\Lambda$CDM, we trade the full high-$\ell$ likelihood for the much faster ``lite'' version and make use of {\sc MultiNest} \cite{Feroz:2008xx} (with 500 livepoints and an evidence tolerance of 0.2) to compute the bayesian evidence. We checked that this gives results which are fully consistent with the MCMC on the full likelihood. We perform model comparison by calculating $\Delta{\rm log} B = {\rm log} B({\rm EDE})-{\rm log} B(\Lambda{\rm CDM})$. Interestingly, we find ``definite'' (or ``positive'') evidence in favor of the EDE cosmology in the $n=3$ and $n=\infty$ model according to the modified Jeffreys' scale \cite{Jeffreys61,Nesseris:2012cq}. While $n=2$ has a better $\chi^2$ than the $n=\infty$ model, it has a weaker evidence. We attribute this to the fact that $n=2$ effectively has one more free parameter since $c_s^2$ depends on $\phi_i$, while $c_s^2=1$ in the $n=\infty$ model.

\begin{figure}[!h]
\centering
\includegraphics[scale=0.5]{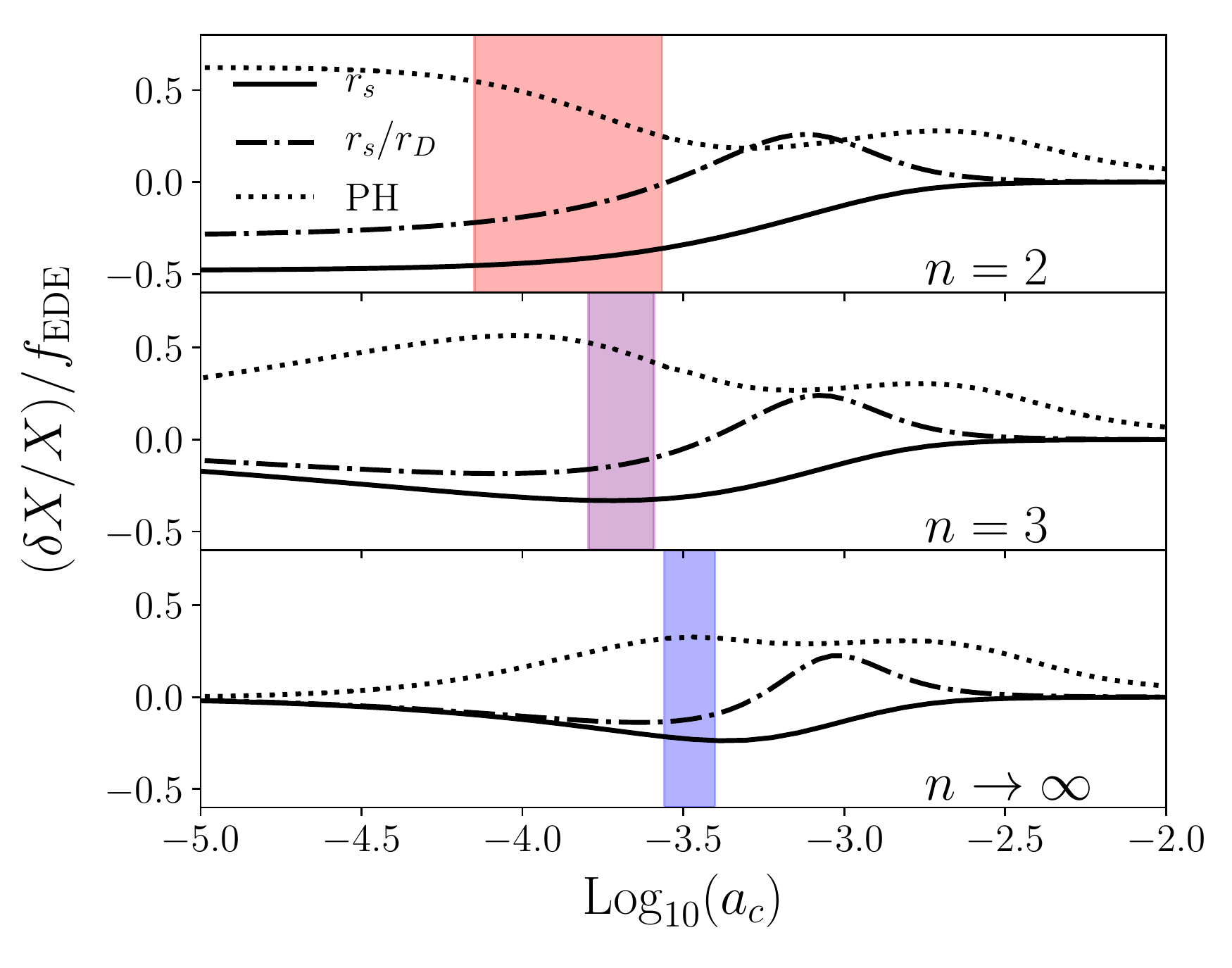}
\caption{The variation of the scales that are `fixed' by the CMB data with respect to $f_{\rm EDE}(a_c)$ as a function of $a_c$ with all other cosmological parameters fixed at their {\it Planck} best-fit values \cite{Aghanim:2018eyx}. The colored bands indicate the marginalized 1$\sigma$ range of $a_c$ for each EDE model considered here.  \label{fig:derivs}}
\end{figure}

One of the most interesting aspects of the EDE resolution of the Hubble tension is that the posterior distributions show that the field must become dynamical 
around matter-radiation equality.  Within the context of $\Lambda$CDM, a simplified picture of the CMB power spectrum can be described by three angular scales: $\ell_{\rm eq}$ (the projected Hubble horizon at matter-radiation equality), $\ell_s$ (the projected photon-baryon sound-horizon at decoupling), and $\ell_D$ (the projected Silk damping scale at decoupling) \cite{Hu:2000ti}.  
These angular scales are given by the ratio of a physical scale at decoupling with the angular diameter distance to the surface of last scattering: $\ell_X = \pi D_A(z_*)/r_X(z_*)$. 
Additionally, the overall amplitudes of the CMB peaks (in particular, the first one) are accurately measured by {\em Planck}. 
It is straightforward to show that ${\rm PH} \propto \omega_{\rm cdm}^{-0.5}$, $\ell_{\rm eq} \propto \omega_{\rm cdm}^{0.5}h^{-0.2}$, $\ell_s \propto \omega_{\rm cdm}^{-0.16} h^{-0.2}$, $\ell_s/\ell_D = r_s/r_D \propto \omega_{\rm cdm}^{0.03}$,
where PH stands for the height of the first peak and we assume that the heights of the even and odd peaks fixes $\omega_b$. 
In  $\Lambda$CDM, the measured peak height determines $\omega_{\rm cdm}$, allowing an inference of $h$ through $\ell_{\rm eq}$, $\ell_s$, and $\ell_D$. Alternatively, 
using the determination of $H_0$ from SH0ES, one would deduce values of $\ell_{\rm eq}$, $\ell_s$, and $\ell_D$ too small compared to their measured values. As shown by several recent studies \cite{Bernal:2016gxb,Evslin:2017qdn,Aylor:2018drw}, this can be re-cast as a mismatch between the sound horizon deduced from {\em Planck} data, and that reconstructed from the standard distance ladder. The value of $r_s$ measured by Planck is higher by $\sim 10$ Mpc compared to that directly deduced from the distance ladder. 

The role of the EDE is to decrease $r_s$, while keeping the angular scales and peak heights fixed  via small shifts in other cosmological parameters. 
For each value of $n$, we show the fractional change in $r_s$, $r_s/r_D$ and PH with $f_{\rm EDE}(a_c)$ as a function of $a_c$ in Fig.~\ref{fig:derivs}. 
The $1\sigma$ errors on $a_c$, reconstructed from our analysis, are also shown. 
Unsurprisingly we find that the value of $a_c$ is driven to be close to the maximal fractional change in $r_s$ (solid line). 
Additionally, one can see that such an EDE leads to a shift in the ratio $r_s/r_D$ (dash-dotted line) and increase in peak height (dotted line). From the above scaling relations it is clear that the increase in the peak height can be compensated by an increase in $\omega_{\rm cdm}$, giving the positive correlation between $f_{\rm EDE}(a_c)$ and $\omega_{\rm cdm}$ visible in the 2D-posterior distribution shown in Fig.~\ref{fig:posteriors}. Moreover, the dynamics of the EDE compensate for such a change in $\omega_{\rm cdm}$, leaving the imprint of $\ell_{\rm eq}$ on the power spectra relatively unchanged.   An increase in $\omega_{\rm cdm}$ leaves $r_s/r_D$ roughly unaffected but this ratio cannot be kept fully fixed. This brings us to our main conclusion: the favored EDE model is the one that, while maximizing the decrease in $r_s$, minimizes\footnote{In practice, a relatively small shift in $r_s/r_D$ is allowed as long as a small shift in $n_s$ can compensate for it, leading to a mild shift in the best-fit value of $n_s$ (see Table~\ref{table:param_values}).} the change in $r_s/r_D$.  
Using these scaling laws, for $n=3$ a resolution of the Hubble tension will roughly require  $\delta\omega_{\rm cdm} \simeq 0.01$ and $f_{\rm EDE}(a_c) \simeq 0.1$ at ${\rm Log}_{10}(a_c) \simeq -3.7$. Strikingly, this crude estimate agrees well with the best-fit values in Table \ref{table:param_values}. This analysis also explains why $n=3$ is favored over the $n=2$ and $n\to\infty$ case.
Moreover, we can understand why the EDE cosmology is a ``better'' resolution of the Hubble tension than increasing the effective number $N_{\rm eff}$ of neutrino degrees of freedom: the effects of an additional radiation energy density can be read off of Fig.~\ref{fig:derivs} for the $n=2$ case at ${\rm Log}_{10}(a_c) \ll -4.5$. In that case, the EDE simply behaves like additional radiation all relevant times. One can see that $r_s/r_D$ is significantly affected, leading to additional tension with the data, as previously noted in Ref.~\cite{Hou:2011ec}.  

We find that it is essential to consistently include perturbations in the EDE fluid. Neglecting perturbations is inconsistent with the requirement of overall energy conservation and therefore leads to unphysical features in the CMB power spectra which restrict the success of the resolution. This, in part, explains why a former study \cite{Karwal:2016vyq} did not find a good fit to the CMB for $f_{\rm EDE}(a_c \simeq 10^{-3.5}) \sim 5\%$.

\begin{figure}[h]
    \centering
    \includegraphics[scale=0.45]{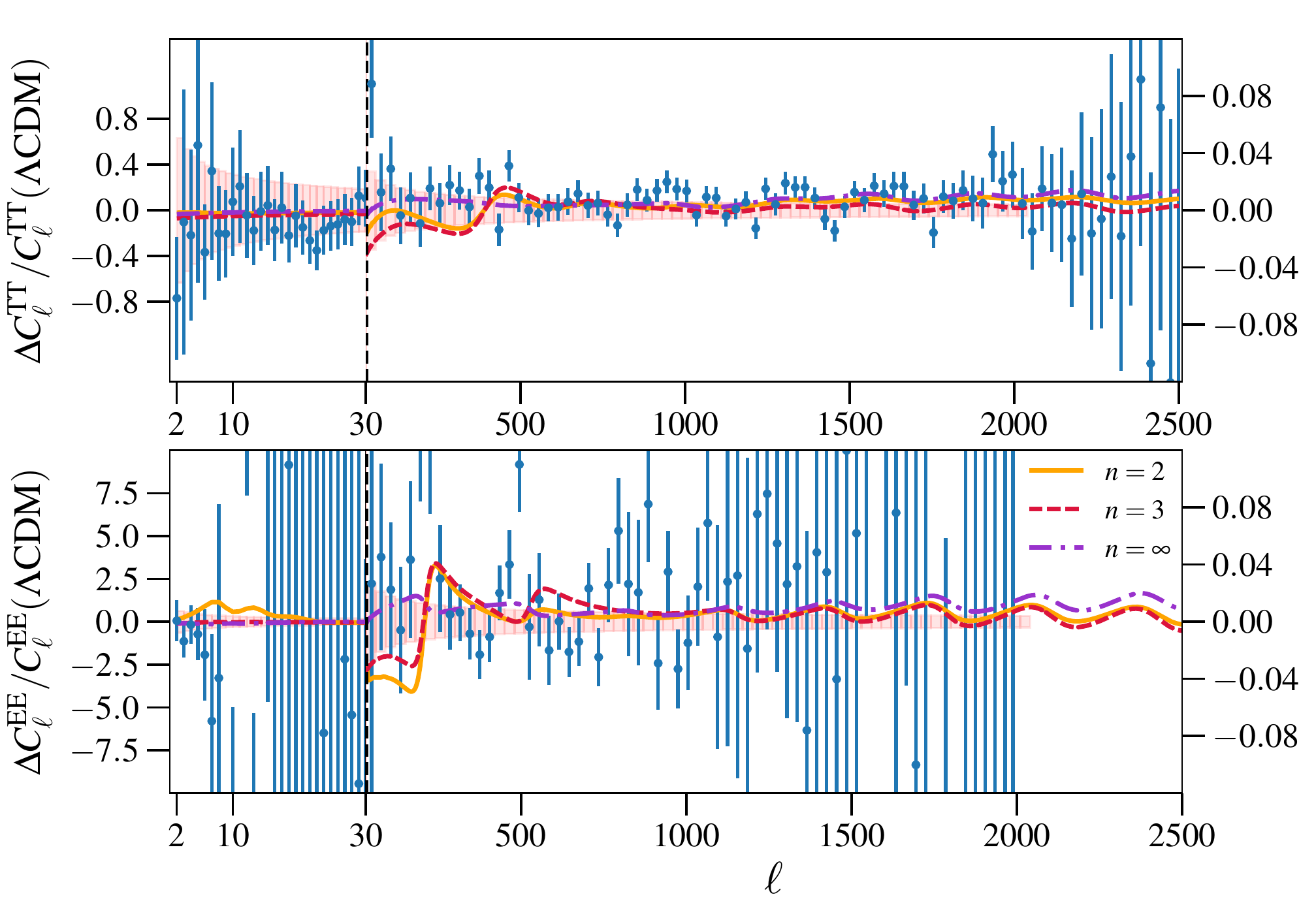}
    \caption{Residuals of the CMB TT (top panel) and EE  (bottom panel) power spectra calculated in the best-fit EDE model with respect to \LCDM, obtained from our MCMC analyses.  Blue points show residuals of {\em Planck} data, while orange bands show the binned Cosmic Variance with the same bins and weights as {\em Planck}.\label{fig:bestfit}}
\end{figure}

In Fig.~\ref{fig:bestfit}, we show the residuals of the CMB TT (top panel) and EE (bottom panel) power spectra calculated in the best-fit EDE model with respect to our best-fit \LCDM\ (i.e.~fit on all datasets).  One can see that the EDE leads to residual oscillations particularly visible at small scales in the EE power-spectrum, which represent an interesting target for next-generation experiments such as the Simons Observatory \cite{Ade:2018sbj}, CMB-S4 \cite{Abazajian:2016yjj} or CoRE \cite{DiValentino:2016foa}. Additionally, the pattern around the first peak ($\ell \sim 30-500$) in the EE spectrum might be detectable in the future by large-scales E-mode measurements such as CLASS \cite{2014SPIE.9153E..1IE} or LiteBird \cite{Suzuki:2018cuy}. Finally, the changes in $r_s$, $n_s$, and $A_s$ leave signatures in the matter power spectrum that can potentially be probed by surveys such as KiDS, DES and Euclid. This can also be seen in the parameter $S_8\equiv\sigma_8(\Omega_m/0.3)^{0.5}$, which is shifted by about 1$\sigma$ upwards from its $\Lambda$CDM value. This slightly increases the so-called ``$S_8$ tension'' (e.g.~\cite{Raveri:2018wln}) and therefore deserves more attention in future work. For example, the tension with the most recent KiDS cosmic-shear measurement \cite{Hildebrandt:2018yau} increases from 2.3$\sigma$ to 2.5$\sigma$.
As a first check, we have performed additional runs including SDSS DR7 \cite{Reid:2009xm} and KiDS \cite{Kohlinger:2017sxk} likelihoods, and found that our conclusions are unaffected.

In this {\it Letter}, we have shown that an EDE that begins to dilute faster than matter at a redshift $z_c \gtrsim 3000$ can explain the increasingly significant (currently $3.8\sigma$) tension between $H_0$ inferred from the CMB \citep{Aghanim:2018eyx} and Cepheid variables/supernovae at low redshifts \citep{Riess:2018byc}.  Using {\em Planck}, BAO measurements, the Pantheon supernovae data, the local SH0ES measurement of $H_0$ and a MCMC analysis, we found that a field accounting for $\sim5\%$ of the total energy density around $z\sim 5000$ and diluting faster than radiation afterwards can solve the Hubble tension without upsetting the fit to other data sets.
We found that in the EDE cosmology the best-fit $\chi^2$ is reduced by $-9$ to $-14$ (with a slight preference for $n=3$) compared to \LCDM\ using the same data-sets. Moreover, the \LCDM\ fit to just the {\em Planck} data is as good as the combined fit to all of the data sets in the EDE cosmology. This is in stark contrast with the popular increased-$N_{\textrm{eff}}$ resolution.

The oscillating field EDE may naturally arise in the `string-axiverse' scenario \citep{Svrcek:2006yi,Arvanitaki:2009fg,Cicoli:2012sz,Kamionkowski:2014zda,Stott:2017hvl}. The standard axion potential is obtained for $n=1$, while higher-$n$ potentials may be generated by higher-order instanton corrections \citep{Kappl:2015esy}.  The EDE resolution of the Hubble tension, along with the current accelerated expansion and the evidence for early-Universe inflation (and perhaps the accelerated expansion postulated \citep{Hill:2018lfx,Poulin:2018dzj} to account for EDGES \citep{Bowman:2018yin}) may suggest that the Universe undergoes episodic periods of anomolous expansion, as suggested in Refs.~\citep{Dodelson:2001fq,Griest:2002cu,Linder:2010wp,Kamionkowski:2014zda,Emami:2016mrt,Karwal:2016vyq}.

A future cosmic-variance-limited experiment around $\ell\sim30-500$ and above $\ell\sim1500$ could probe the specific residual oscillations in the CMB power spectra associated with the EDE dynamics, while the shifts in $A_s$, $n_s$, $r_s$, and $k_{\rm eq}$ will be probed by future LSS surveys. 

{\em Note added:} as this work was being completed, a new value of $H_0$ was published by SH0ES increasing the tension with \LCDM~from {\em Planck} to $4.4\sigma$ \cite{Riess:2019cxk}.

We thank Adam Riess and Tommi Tenkanen for useful conversations. We thank Antony Lewis and Colin Hill for detailed comments. This research was conducted using computational resources at the Maryland Advanced Research Computing
Center (MARCC) and the Universit\'{e} Savoie Mont-Blanc MUST computing center. This work was supported at JHU by NSF Grant No.\ 1818899, NASA NNX17AK38G, and the Simons Foundation.
TLS acknowledges support from NASA 80NSSC18K0728 and the Hungerford Fund at Swarthmore College.

\begin{table*}[th]
  \begin{tabular}{|l|c|c|c|c|c|}
    \hline\hline
    Datasets &~~$\Lambda$CDM~~&~~~$n=2$~~~&~~~ $n=3$~~~&~~~$n=\infty$~~~&~~~$N_{\rm eff}$~~~\\ \hline \hline
    \textit{Planck} high-$\ell$ & 2449.5 & 2448.4 & 2445.9 & 2445.4 &  2451.9\\
    \textit{Planck} low-$\ell$ & 10494.7 & 10494.2  & 10492.8 & 10493.8 & 10493.8\\
    \textit{Planck} lensing& 9.2 & 9.4 & 9.6 & 11.7 & 9.8 \\
    BAO-low $z$ & 1.7 & 2.1 & 2.1 & 1.8 & 2.7 \\
    BAO-high $z$& 1.8 & 1.9 & 1.9 & 1.9 & 2.0\\
    Pantheon & 1027.1  & 1027.3 &  1026.9 & 1026.9& 1027.1 \\
    SH0ES& 11.1 & 2.3 & 1.4 & 4.6 & 3.9\\
    \hline
    Total $\chi^2_\mathrm{min}$   & 13995.1  &  13985.6 & 13980.6 & 13986.0 & 13991.2 \\
    $\Delta \chi^2_\mathrm{min}$  & 0 & -9.5 & -14.5 & -9.1 & -3.9   \\ 
    $\Delta\log B\footnote{The evidence has been calculated from the ``lite'' version of the high-$\ell$ likelihood.}$  & 0 & -0.51 & +2.51 & +2.41 & -0.44 \\ 
    \hline
  \end{tabular}
  \caption{The best-fit $\chi^2$ per experiment for the standard $\Lambda$CDM model, the EDE cosmologies and \LCDM+$N_{\rm eff}$. The BAO-low $z$ and high $z$ datasets correspond to $z\sim 0.1-0.15 $ and $z \sim 0.4-0.6$, respectively.  For comparison, using the same {\sc CLASS} precision parameters and {\sc MontePython}, a $\Lambda$CDM fit to {\it Planck} data only yields $\chi^2_{{\rm high}-\ell}\simeq2446.2$, $\chi^2_{{\rm low}-\ell}\simeq10495.9$ and $\chi^2_{{\rm lensing}}\simeq9.4$ with $R-1<0.008$.}
  \label{table:chi2_preliminary}
\end{table*}

\begin{table*}
  \begin{tabular}{|l|c|c|c|c|}
    \hline\hline1.04149
    Parameter &~~$\Lambda$CDM~~&~~~$n=2$~~~&~~~ $n=3$~~~&~~~$n=\infty$~~~\\ \hline \hline
    $100~\theta_s$ & $1.04198
    ~(1.04213)\pm0.0003$ & $1.04175~(1.0414)_{-0.00064}^{+0.00046}$ & $1.04138~(1.0414)\pm0.0004$ & $1.04159~(1.04149)\pm0.00035$\\
    $100~\omega_b$ & $2.238~(2.239)\pm 0.014$ & $2.244~(2.228)_{-0.022}^{+0.019}$ &$2.255~(0.258)\pm0.022$ & $2.257~(2.277)\pm0.024$ \\
    $\omega_{\rm cdm}$&$0.1179~(0.1177)\pm0.0012$& $0.1248~(0.1281)_{-0.0041}^{+0.003}$ & $0.1272~(0.1299)_\pm0.0045$ &
    $0.1248~(0.1249)\pm0.0041$\\
    $10^{9}A_s$& $2.176~(2.14)\pm0.051$ &$2.185~(2.230)\pm0.056$ & $2.176~(2.177)\pm0.054$& $2.151~(2.177)\pm0.051$ \\
    $n_s$& $0.9686~(0.9687)\pm 0.0044$ &$0.9768~(0.9828)_{-0.0072}^{+0.0065}$ &$0.9812~(0.9880)\pm0.0080$ & $0.9764~(0.9795)\pm0.0073$ \\
    $\tau_{\rm reio}$ &  $0.075~(0.068)\pm 0.013$& $0.075~(0.083)\pm0.013$& $0.068~(0.068)\pm0.013$ &$0.062~(0.066)\pm0.014$\\
    ${\rm Log}_{10}(a_c)$& $-$  &$-4.136~(-3.728)_{-0.013}^{+0.57}$ &$-3.737~(-3.696)_{-0.094}^{+0.110}$  &   $-3.449~(-3.509)_{-0.11}^{+0.047}$\\
    $f_{\rm EDE}(a_c)$  &$-$  & $0.028~(0.044)_{-0.016}^{+0.011}$& $0.050~(0.058)_{-0.019}^{+0.024}$& $0.054~(0.057)_{-0.027}^{+0.031}$ \\
    \hline
    $r_s(z_{\rm rec})$ & $145.05~(145.1)\pm0.26$ & $141.4~(139.8)_{-1.5}^{+2}$&  $140.3~(138.9)_{-2.3}^{+1.9}$ & $141.6~(141.3)_{-2.1}^{+1.8}$\\
    $S_8$ & $0.824~(0.814)\pm0.012$&$0.826~(0.836)\pm0.014$  & $0.838~(0.842)\pm 0.015$ & $0.836~(0.839)\pm0.015$\\
    $H_0$  & $68.18~(68.33)\pm 0.54$ &$70.3~(71.1)\pm1.2$ &$70.6~(71.6)\pm1.3$&  $69.9~(70)\pm1.1$ \\
    \hline
  \end{tabular}
  \caption{The mean (best-fit) $\pm1\sigma$ error of the cosmological parameters reconstructed from our combined analysis in each model.}
  \label{table:param_values}
\end{table*}

\bibliography{ULA_H0.bib}

\end{document}